\documentclass[runningheads]{llncs}
\usepackage[T1]{fontenc}
\usepackage{graphicx}
\usepackage{booktabs}

\usepackage{biblatex} 
\usepackage{xcolor}
\newcommand\todo[1]{\textcolor{red}{TODO}}
\usepackage{mathtools}
\usepackage{amsmath}
\usepackage{hyperref}
\usepackage{amsfonts}
\usepackage{color}

\usepackage{algorithm}
\usepackage{algorithmic}
\usepackage{listings}

\usepackage{graphicx, subcaption}
\usepackage{colortbl} 
\usepackage{multirow}
\usepackage{multicol}
\usepackage{caption}
\usepackage[utf8]{inputenc}
\usepackage[T1]{fontenc}
\usepackage{enumitem}

\usepackage{array, booktabs, ragged2e}

\begin{document}

\title{A Proposal for Uncovering Hidden Social Bots via Genetic Similarity}
\titlerunning{A Proposal for Uncovering Hidden Social Bots via Genetic Similarity}
%
\author{Edoardo Allegrini\inst{1}\orcidID{0009-0003-8842-6873} \and
Edoardo Di Paolo\inst{1}\orcidID{0000-0001-9216-8430} \and
Marinella Petrocchi\inst{2,3}\orcidID{0000-0003-0591-877X}
\and
Angelo Spognardi\inst{1}\orcidID{0000-0001-6935-0701}}
\authorrunning{Edoardo Allegrini et al.}
%
\institute{Computer Science Department, Sapienza University of Rome, Italy \email{allegrini@di.uniroma1.it,dipaolo@di.uniroma1.it,spognardi@di.uniroma1.it}\\ \and
Istituto di Informatica e Telematica, CNR, Pisa, Italy
\email{marinella.petrocchi@iit.cnr.it}\\
 \and
Scuola IMT Alti Studi Lucca, Italy\\
}
\maketitle              
\begin{abstract}
Social media platforms face an ongoing challenge in combating the proliferation of social bots, automated accounts that are also known to distort public opinion and support the spread of disinformation. 
Over the years, social bots have evolved greatly, often becoming indistinguishable from real users, and more recently, families of bots have been identified that are powered by Large Language Models to produce content for posting. 
We suggest an idea to classify social users as bots or not using genetic similarity algorithms. These algorithms provide an adaptive method for analyzing user behavior, allowing for the continuous evolution of detection criteria in response to the ever-changing tactics of social bots. Our proposal involves an initial clustering of social users into distinct macro species based on the similarities of their timelines. Macro species are then classified as either bot or genuine based on genetic characteristics.
The preliminary idea we present, once fully developed, will allow existing detection applications based on timeline equality alone to be extended to detect bots. By incorporating new metrics, our approach will systematically classify non-trivial accounts into appropriate categories, effectively peeling back layers to reveal non-obvious species.

\end{abstract}

\section{Introduction}
    
The digital age has brought with it an unprecedented proliferation of accounts on social platforms, resulting in a diverse and complex ecosystem. Of particular note are automated accounts, commonly known as social bots \cite{ferrara2016rise,cresci2020decade}. These digital artifacts have received increasing attention, not only because of their ubiquity, but also because of the role they often play as vehicles for misinformation and propaganda \cite{shao2018spread}.
Recently, families of bots have been heuristically identified that use
large language models to produce content for
 publication. The proliferation of these advanced bots raises concerns about the inability of researchers to detect them \cite{fox8}. 

One of the most important strands of research began with the realization that bots, programmed to pursue specific goals, often operate in a coordinated manner and exhibit similar behaviors. In particular, one modeling and detection technology that was notably relevant was that based on digital DNA \cite{DBLP:journals/expert/CresciPPST16,dna_pres}. 
Digital DNA is a string of characters, each of which associated with a specific account action, representing the timeline of the account. This modeling technique has been used in several studies, see, e.g.,~\cite{dipaolo2023dna,pasricha2019detecting,gilmary2023entropy,chawla2023hybrid}. 

Based on the intuition that accounts of the same type behave similarly, if not exactly the same, we  
propose a method for classifying bot accounts that may be mistaken for real accounts.

\section{Proposed Method}
    \begin{figure}[h!]
        \centering
        \includegraphics[width=.9\textwidth]{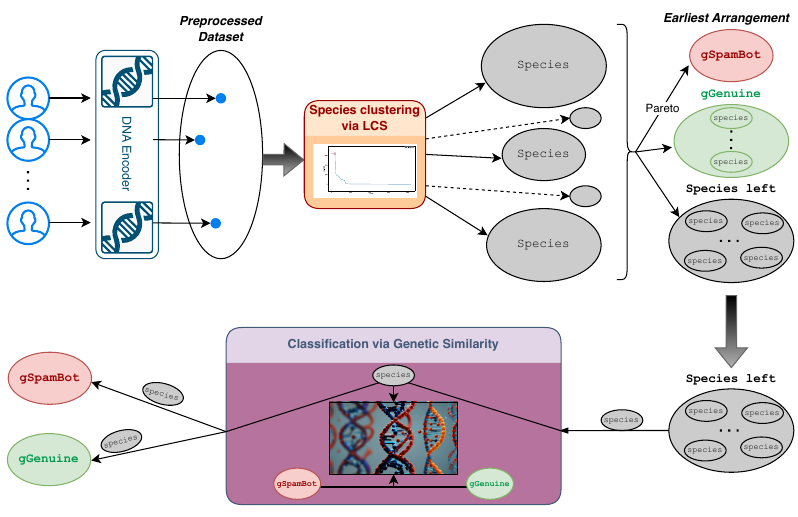}
        \caption{
        Steps for hidden social bots classification\label{fig:classificationProcess}}
    \end{figure}

    The classification approach we propose has been designed for Twitter/X users, but minor adjustments can easily adapt it for use on other social media platforms. \autoref{fig:classificationProcess} shows the scheme of the procedure.
    
    \paragraph{Digital DNA}
        The first phase involves pre-processing users by encoding their online behavior using Digital DNA. Digital DNA aims to compactly represent the behavior of a social account, using a sequence of characters from an alphabet $\mathbb{B}$, such as the following made of three characters:
\begin{equation}
    \label{eq:alphATC}
    \mathbb{B}^{3}_{\textit{type}} = \begin{Bmatrix}
    A \rightarrow \text{plain tweet}\\ 
    T \rightarrow \text{retweet} \\ 
    C \rightarrow \text{reply}
    \end{Bmatrix}
\end{equation}

\begin{figure}[htb]
    \centering
    \includegraphics[width=.7\textwidth]{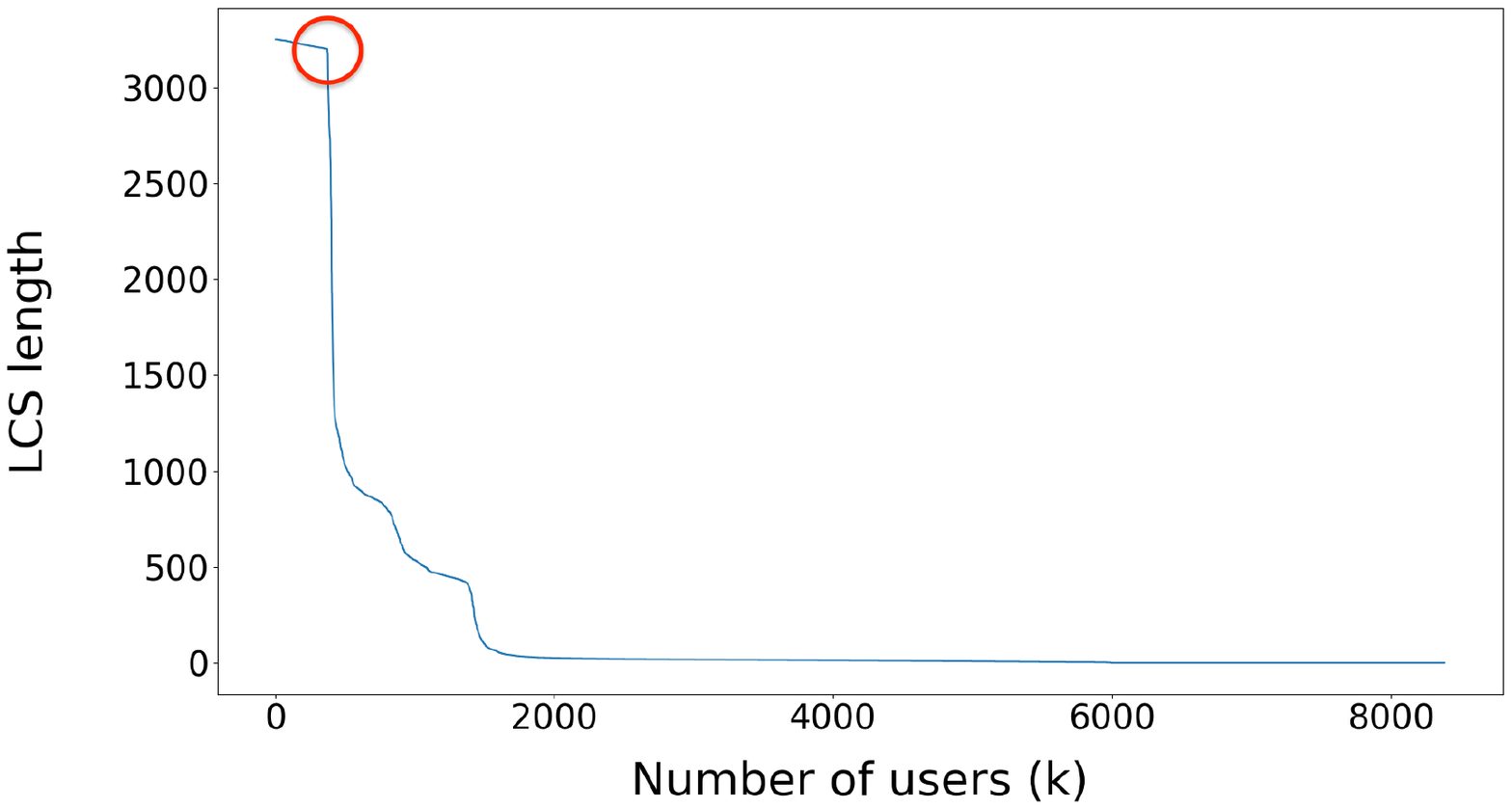}
    \caption{Plot of an illustrative LCS curve 
    \label{fig:function_LCS_C17}}
\end{figure}

    \paragraph{Clustering users into species}
        After the pre-processing stage, users are grouped into macro species, based on the concept of the Longest Common Substring (LCS) \cite{lcs2011}. 
The LCS of two or more strings is the longest string that is a substring of all of them.
Figure \ref{fig:function_LCS_C17} shows an example of an LCS curve, where the abscissae are the groups of k users, and the ordinate is the length of the LCS for each of the user groups.
Considering a dataset of $N$ users paired with their digital DNA sequences, the LCS computation is performed in linear time \cite{lcs2011} between $k$ users, where $k \in \{2,...,N\}$. 
LCS is an indicator of behavioral similarity within the user group:
When the sub-DNAs in the LCS curve are of approximately constant length, we can deduce that the users with these sequences have similar behavior. Conversely, if we observe a significant drop in the LCS curve, we know that the behavior of newly added users differs considerably from that of the users in the previous group. 
As an example, the red circle in the figure marks the point where the curve exhibits a significant drop. 
To identify and cluster users into species, the first significant drop in the LCS curve is detected, which marks a behavioral shift. Users associated with that drop are grouped into a new species and subsequently removed. The LCS curve is then recomputed for the remaining users that have not been assigned to a cluster yet. This process is repeated, progressively segmenting and grouping all users in the dataset into distinct clusters called \textit{species}.

    \paragraph{Spambot and genuine accounts: first arrangement}
        From the macro clusters obtained, two key groups are constructed: \textit{gSpamBot}, by selecting among the species the one demonstrating evident bot-like behavior, and \textit{gGenuine}, with a high predominance of genuine users. 
Later, these two groups will be populated by the remaining users, by adopting specific algorithms developed to measure the genetic similarity.

To establish the initial \textit{gSpamBot} and \textit{gGenuine} groups, the following idea is applied: the LCS of a species represents the users within it; therefore, a group with a long LCS indicates similar social behavior, possibly an indicator of social bots. On the other side, a group with a short LCS implies a diversity in the social behavior of its members. The construction of the initial \textit{gSpamBot} group is inspired by the Pareto principle, which seeks to determine the subset of individuals that have the most significant impact on the overall community. Based on the original dataset, users have now been categorized into three groups: \textit{gSpamBot}, which comprises a considerable number of users exhibiting very similar behavior, strongly indicating that they are social bots; \textit{gGenuine}, formed by merging species that demonstrate human-like actions; and lastly, species that do not fall into either \textit{gSpamBot} or \textit{gGenuine} and are therefore unlabeled at this stage.

    \paragraph{Classification of species using genetic similarity}
        To classify the unlabeled species (those colored gray in \autoref{fig:classificationProcess}), whose users are not immediately classifiable as bots or not, we propose an algorithm that uses custom genetic similarity metrics. The idea is using a sequence alignment algorithm —a well-established technique in bioinformatics— to compare the LCS of each unlabeled species with those of the two primary groups. After this alignment, our algorithm introduces a structured classification process. The process involves two key steps: first, calculating a similarity score based on the alignment of LCS sequences between species; and second, evaluating a new metric that considers the relative similarity of DNA sequences within a species and the size of the population contributing to that similarity. By integrating these procedures, we believe that our approach can effectively label the previously unlabeled species as either \textit{gSpamBot} or \textit{gGenuine}, thereby completing the account classification process.

\section{Conclusions}

    In this short paper, we presented the idea of a new approach to social bot detection that we hope will not only achieve effective classification, but also maintain a transparent decision-making process. 

We plan to implement and test our proposed classifier using well-established bot repositories (see the ones published on the \href{https://botometer.osome.iu.edu/bot-repository/}{site} of the OSOME research unit at Indiana University) as well as recently discovered datasets where social bots use Large Language Models (LLMs) to write their posts.
    
\begin{credits}
\subsubsection{\ackname} This work is partially supported by project SERICS (PE00000014) under the NRRP MUR program funded by the EU - NGEU; 
by project re-DESIRE (DissEmination of ScIentific REsults 2.0), funded by IIT-CNR; by project `Prebunking: predicting and mitigating coordinated inauthentic behaviors in social media', funded by Sapienza University of Rome.

\end{credits}
%
%
%
%
\printbibliography
\end{document}